\documentclass[preprint,prl,superscriptaddress,prbib,11pt]{revtex4-1}
\usepackage{graphicx}
\usepackage{bm}
\usepackage{hyperref}
\usepackage{todonotes}
\usepackage{nicefrac}
\usepackage{bbm}
\usepackage{bbm}
\usepackage{amsmath}
\usepackage[normalem]{ulem}
\usepackage{booktabs}
\usepackage{subfigure}
\usepackage{rotating}
\usepackage{ragged2e}
\usepackage{color}
\usepackage{float}
\usepackage[T1]{fontenc}
\usepackage[utf8]{inputenc}
\patchcmd{\subsubsection}{\itshape}{\bfseries}{}{}
\usepackage{caption}
\usepackage{soul}
\usepackage{geometry}
\begin{document}

\title{Discovery of an odd-parity f-wave charge order in a kagome metal}

\author{Jiangchang Zheng}
\thanks{These authors contributed equally.}
\affiliation{Department of Physics, The Hong Kong University of Science and Technology, Clear Water Bay, Kowloon, Hong Kong SAR}

\author{Caiyun Chen}
\thanks{These authors contributed equally.}
\affiliation{Department of Physics, The Hong Kong University of Science and Technology, Clear Water Bay, Kowloon, Hong Kong SAR}

\author{Ruiqin Fu}
\thanks{These authors contributed equally.}
\affiliation{Institute of Theoretical Physics (ITP), Chinese Academy of Sciences, Beijing 100190, China}

\author{Luca Buiarelli}
\affiliation{Chemical Engineering and Materials Science, University of Minnesota, Minneapolis, Minnesota 55455, USA}

\author{Zihan Lin}
\affiliation{Department of Physics, The City University of Hong Kong, Kowloon, Hong Kong SAR}

\author{Fazhi Yang}
\affiliation{Department of Physics, The City University of Hong Kong, Kowloon, Hong Kong SAR}

\author{Tianhao Guo}
\affiliation{Department of Physics, The City University of Hong Kong, Kowloon, Hong Kong SAR}

\author{Ganesh Pokharel}
\affiliation{Materials Department, University of California, Santa Barbara, Santa Barbara, California 93106, USA}
\affiliation{Department of Natural Sciences, University of West Georgia, Carrollton, GA 30118, USA}

\author{Andrea Capa Salinas}
\affiliation{Materials Department, University of California, Santa Barbara, Santa Barbara, California 93106, USA}

\author{Sen Zhou}
\affiliation{Institute of Theoretical Physics (ITP), Chinese Academy of Sciences, Beijing 100190, China}

\author{Turan Birol}
\affiliation{Chemical Engineering and Materials Science, University of Minnesota, Minneapolis, Minnesota 55455, USA}

\author{Stephen D. Wilson}
\affiliation{Materials Department, University of California, Santa Barbara, Santa Barbara, California 93106, USA}

\author{Junzhang Ma}
\affiliation{Department of Physics, The City University of Hong Kong, Kowloon, Hong Kong SAR}

\author{Daniel J. Schultz}
\email{daniel.schultz@kit.edu}
\affiliation{Institute for Theoretical Condensed Matter Physics, Karlsruhe Institute of Technology, Wolfgang Gaede Strasse 1, 76131 Karlsruhe, Germany}

\author{Xianxin Wu}
\email{xxwu@itp.ac.cn}
\affiliation{Institute of Theoretical Physics (ITP), Chinese Academy of Sciences, Beijing 100190, China}

\author{Berthold J\"ack}
\email{bjaeck@ust.hk}
\affiliation{Department of Physics, The Hong Kong University of Science and Technology, Clear Water Bay, Kowloon, Hong Kong SAR}

\date{\today}

\begin{abstract}
The spontaneous breaking of symmetries is a cornerstone of physics, defining the phases of matter from the cosmological scale to the quantum realm. In condensed matter, electronic orders are classified by their behavior under fundamental symmetries like spatial inversion (parity). While even-parity orders, such as conventional superconductivity and charge density waves, are ubiquitous, their odd-parity counterparts--predicted to host exotic phenomena such as gapless quasiparticle excitations and novel collective modes--are comparatively elusive states of quantum matter. Here, using high-resolution scanning tunneling microscopy and angle-resolved photoemission spectroscopy on the kagome metal CsV$_3$Sb$_5$, we report the discovery of an inversion symmetry-breaking $f$-wave charge bond order. We show that this phase, which preserves translation symmetry, is stabilized by the spontaneous opening of a spectral gap at a previously overlooked Dirac point, providing a textbook condensed-matter realization of the Gross-Neveu model for dynamical mass generation and parity breaking. Intriguingly, this $f$-wave order is itself a intervening phase, vanishing abruptly below a temperature of 10\,K and pointing to a subsequent transition into a `hidden' electronic state that is invisible to local STM probes. Our findings establish odd-parity charge order as a novel phase of matter, here, embedded within the intricate hierarchy of correlated electronic orders on the kagome lattice.
\end{abstract}

\maketitle

\subsection{Introduction}

The spontaneous breaking of symmetries within a material~\cite{landau1937theory} underpins the modern understanding of condensed matter physics and is a key mechanism for realizing novel quantum phases of matter. When electrons in a material collectively organize into an ordered phase, the resulting order parameter can be classified by its behavior under the symmetries of the underlying crystal lattice. A crucial distinction is its parity: "even-parity" orders remain unchanged under spatial inversion (${\bf k} \rightarrow -{\bf k}$ in momentum space), while "odd-parity" orders change sign. This distinction has profound consequences. Even-parity superconductors, for instance, have a gap that is symmetric in momentum space. In contrast, odd-parity superconductors, like the $p-$wave state, can have nodes in their order parameter and host topological quasiparticle excitations, such as Majorana quasiparticles~\cite{mackenzie2000p, kitaev2001unpaired, sato2010topological, mourik2012signatures, nadj2014observation, jiao2020chiral}. Similarly, systems with long-range magnetic order can be categorized, giving rise to the emerging landscape of $p-$wave magnets~\cite{song2025electrical, yamada2025metallic} and altermagnets~\cite{vsmejkal2022beyond} with odd- and even-parity order parameter, respectively.

This same classification principle applies to charge-ordered states, or charge density waves (CDWs), which arise when electronic interactions cause the charge density to spontaneously modulate. Conventional CDW states break the translational symmetry of the underlying crystal lattice~\cite{frohlich1954theory, chen2016charge}, resulting in a lattice distortion that augments the unit cell. This scenario is illustrated in Fig.~\ref{fig:fig1},~(a) and (b) for a $2a\times2a$ CDW appearing on the kagome lattice (often described as Star-of-David/$3\bf Q$ CDW in the literature) with wave vectors $\boldsymbol{q} = (\frac{\pi}{a},\frac{\pi}{\sqrt{3}a}), (-\frac{\pi}{a},\frac{\pi}{\sqrt{3}a}), (0,\frac{2\pi}{\sqrt{3}a})$, which increases the unit cell area by a factor of four ($a$ denotes the length of the lattice vector). Because the charge order pattern of such a phase only breaks translational symmetry, the associated order parameter is of $s-$wave symmetry, has even parity, and is rotationally invariant in momentum space. However, recent work on kagome lattice and square-net materials suggests that the physics of charge density waves can be substantially richer~\cite{kotliar1988resonating, affleck1988large, hsu1991two, nayak2000density, feng2021chiral, denner2021analysis}, encompassing symmetry-breaking orders and collective mode excitations~\cite{jiang2021unconventional, zhao2021cascade, mielke2022time, xu2022three, wu2022simultaneous, guo2022switchable, khasanov2022time, wang2022axial, gui2025probing, wang2025long, singh2025ferroaxial, zheng2025quasiparticle}. Moreover, in analogy with their superconducting and magnetic counterparts, there exists the possibility of odd-parity charge orders with finite angular momentum ($l=1,3,...$, i.e., $p$- or $f$-wave symmetry). Odd-parity charge order parameters are thus characterized by a nodal momentum-space structure that can support its own class of novel gapless excitations and collective modes~\cite{nayak2000density}. A characteristic feature of these odd-parity charge orders is the rearrangement of charge density within each unit cell, breaking local inversion symmetry. Hence, they can even appear as a ${\bf q}=0$ order that does not break translation symmetry. An example for a ${\bf q}=0$ $f-$wave charge order on the kagome lattice which breaks inversion ($\mathcal{P}$) and mirror-xz ($\mathcal{M}_{xz}$) symmetries, is shown in Fig.~\ref{fig:fig1}(d).

While earlier studies of underdoped cuprates found evidence for a CDW with even parity $d-$wave ($l=2$) structure factor~\cite{fujita2014direct, comin2015symmetry}, experimental evidence for odd-parity charge orders has remained elusive until now. This is for two primary reasons. First, charge orders with ${\bf q}=0$ are inherently fragile. The energetic gains from forming an intra-unit-cell order are typically smaller than for conventional, nesting-driven CDWs, that result in translation-symmetry breaking charge density patterns, because it affects a smaller fraction of electronic states at the Fermi surface~\cite{fernandes2014drives}. Second, their detection is exceptionally challenging, as their defining charge modulation is hidden within the atomic unit cell, requiring probes with high spatial resolution like scanning tunneling microscopy (STM)~\cite{nayak2000density}. Yet, the discovery of parity breaking charge orders provides new platforms for exploring the physics of novel collective modes and their potential connection to other unconventional ground states, such as topological and high-temperature superconductivity~\cite{wheatley1988interlayer, chakravarty1993interlayer, hirsch1989bond}.

Here, we report the experimental discovery of an odd-parity, $f-$wave charge bond order existing at ${\bf q}=0$ on the surface of the kagome metal CsV$_3$Sb$_5$. Using high-resolution spectroscopic mapping with STM, we directly visualize an inversion symmetry-breaking charge density modulation within the kagome unit cell, as shown in Fig.~\ref{fig:fig1}(e). Combining angle-resolved photoemission spectroscopy (ARPES) with model calculations, we show that this order is stabilized by opening a spectral gap at a previously overlooked Dirac point along the $\Gamma-K$ line of the Brillouin zone shown in Fig.~\ref{fig:fig1}(f). Our finding provides a rare and elegant condensed-matter realization of the dynamical generation of mass for Dirac fermions via spontaneous parity symmetry-breaking, as described by Gross and Neveu~\cite{gross1974dynamical}. Intriguingly, we find that this $f-$wave charge bond order is itself a intervening phase; it appears upon cooling but then vanishes abruptly below 10\,K. The disappearance of this order, without any other detectable density of states signature in our scanning tunneling microscopy measurements, strongly points to a subsequent phase transition into a new, electronically distinct ground state—potentially a `hidden' order. Our work thus solves the long-standing puzzle of whether odd-parity charge order can exist, while simultaneously revealing a yet to be discovered connection to the material's low-temperature ground state~\cite{ortiz2019new, chen2021roton, zhong2023nodeless, deng2024chiral}.

\subsection{Discovery of inversion symmetry breaking $q=0$ charge order in Ti-doped CsV$_3$Sb$_5$}

\begin{figure}[h]
    \centering
    \includegraphics[width=1\textwidth]{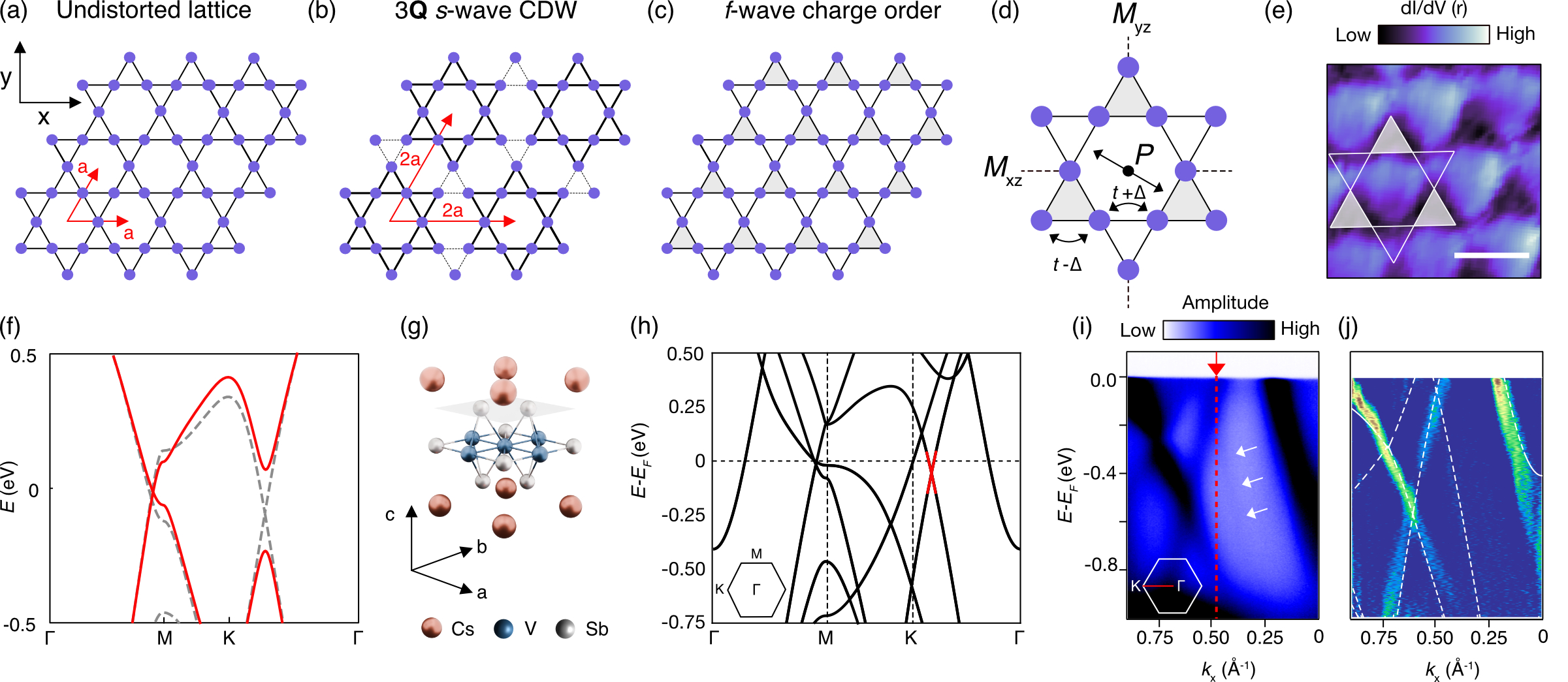}
     \caption{\justifying{\bf Inversion symmetry breaking charge order on the kagome lattice.} (a)-(c), shown are schematic illustrations of the two-dimensional kagome lattice without lattice distortion, in the presence of a conventional $2a\times2a$ CDW with 3-$\mathbf{Q}$ Star-of-David order at the three $M$-points, and in the presence of a an inversion symmetry breaking $f$-wave charge order with ${\bf q}=0$ and $l=3$, respectively. The CDW in panel b leads to a lattice distortion with Star-of-David pattern, augmenting the unit cell area by a factor of four. The ${\bf q}=0$ charge order in panel c retains translation symmetry and modulates the charge density within the unit cell (grey shaded area). (d) The charge density pattern of an $f-$wave charge order with $q=0$ breaks the mirror-$xz$ symmetry $\mathcal{M}_{xz}$ and inversion symmetry $\mathcal{P}$ of the kagome lattice, resulting in an effective three-fold rotation symmetry $C_{3z}$. (e) Shown is a $\mathcal{P}$-breaking real space modulation of the differential tunnel conductance ($dI/dV$) amplitude as measured on the surface of pristine CsV$_3$Sb$_5$ with the STM ($V_{\rm B}=5\,$mV $I=4\,$nA, $V_{\rm m}=1\,$mV, $T=15\,$K, scale bar 5\,Å). (f) Shown is the calculated spectral function of a kagome lattice model without (dashed lines) and with (solid lines) inclusion of an $f-$wave order parameter (${\bf q}=0$, $l=3$), as described in the Methods section. (g) Shown is an isometric view of the atomic lattice structure of the kagome metal CsV$_3$Sb$_5$. (h) Shown is the calculated spectral function of CsV$_3$Sb$_5$ plotted along high symmetry lines within the two-dimensional hexagonal Brillouin zone. (i) and (j) Shown are the spectral function measured on the surface of Cs(V$_{0.95}$Ti$_{0.05}$)$_3$Sb$_5$ at a temperature $T=8\,$K using angle-resolved photoemission spectroscopy at out-of-plane momentum $k_z=0$ and the corresponding second derivative along the $\Gamma-K$ direction, respectively. See Methods section for measurement details. The calculated bands of panel h are overlaid to the second derivative as white dashed lines, and the upward dispersing band is indicated by white arrows in panel i.}
    \label{fig:fig1}
\end{figure}

The kagome metal CsV$_3$Sb$_5$ is a member of the $A$V$_3$Sb$_5$ ($A=$Cs,~K,~Rb) family and has emerged as a primary platform to study symmetry breaking electronic orders favored by the interplay of sublattice interference and charge bond fluctuations~\cite{kiesel2013unconventional, feng2021chiral, denner2021analysis, park2021electronic, fu2025exotic,zhan2026loop}. It crystallizes in a hexagonal structure ($a=b=5.4\,$Å, $c=9\,$Å) and is composed of a vanadium (V) kagome lattice which is coordinated by antimony (Sb) stacked with caesium (Cs) layers along the crystallographic $c$-direction [as seen in Fig.~\ref{fig:fig1}(g)]~\cite{ortiz2019new}. A previous focus in the study of this material has been an even parity $2a\times2a$ CDW [see Fig.~\ref{fig:fig1}(b)]~\cite{jiang2021unconventional, zhao2021cascade, mielke2022time, xu2022three, wu2022simultaneous, guo2022switchable, khasanov2022time, gui2025probing, wang2025long},  which arises from a $3{\bf Q}$-scattering mechanism between the van Hove singularities at the crystallographic $M$-points [as seen in Fig.~\ref{fig:fig1}(h)]~\cite{kang2022twofold, hu2022rich}. In this study, we will focus on a Dirac band crossing along the $\Gamma-K$ line [highlighted in red color in Fig.~\ref{fig:fig1}(h)] that can be detected near the Fermi energy in ARPES measurements~\cite{hu2022rich}, as seen in Fig.~\ref{fig:fig1},~(i) and (j). This spectral feature has received little attention so far and plays a vital role for the formation of the odd-parity $f-$wave charge bond order.

\begin{figure}[H]
    \centering
    \includegraphics[width=1\linewidth]{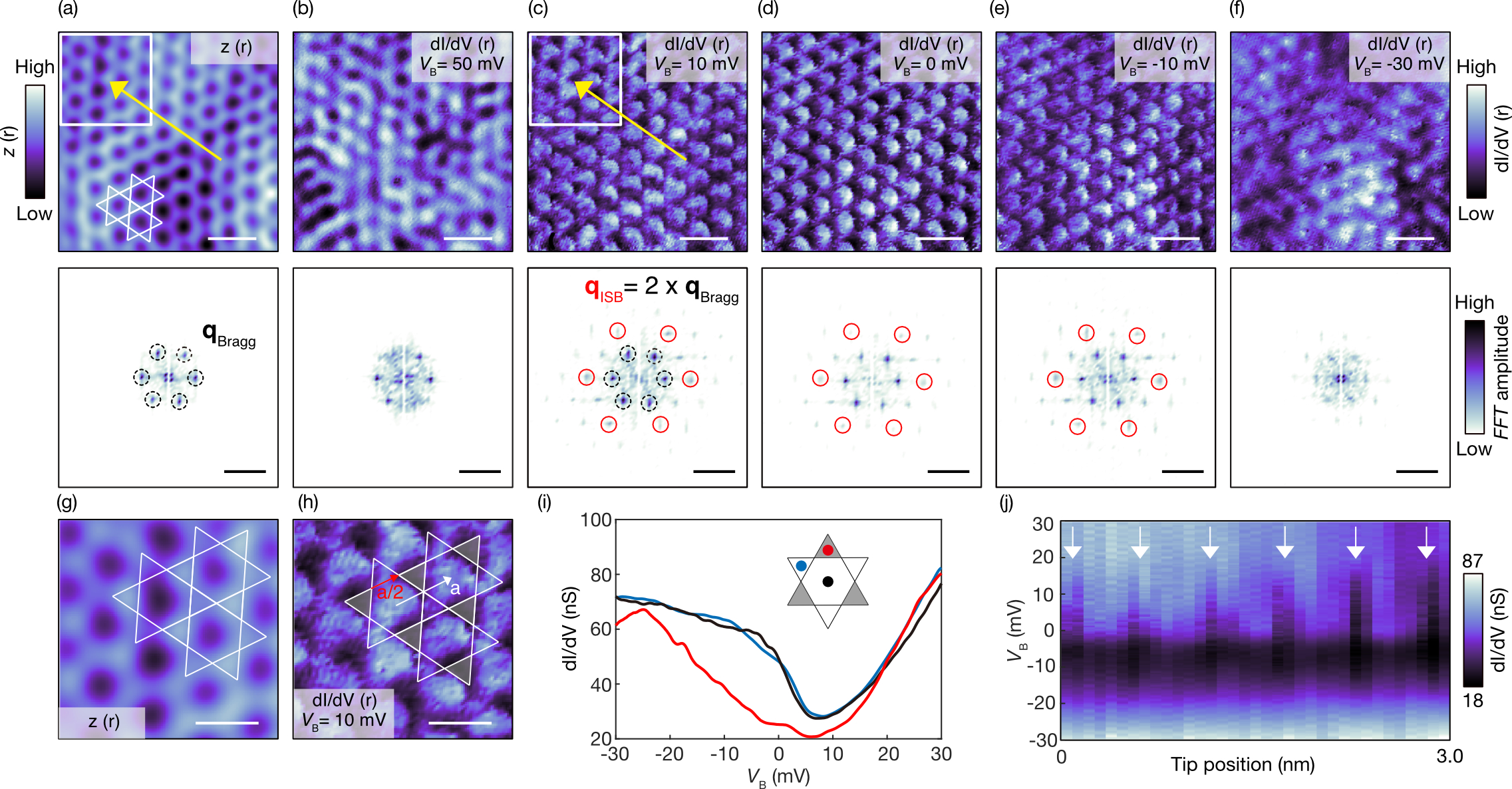}
    \caption{\textbf{Detection of inversion symmetry-breaking density of states at the surface of Cs(V$_{0.95}$Ti$_{0.05}$)$_3$Sb$_5$.} (a) Shown is an STM topography (top panel, $V_{\rm B}=50\,$mV, $I=4\,$nA, $T=14\,$K, scale bar 1\,nm)) and corresponding two-dimensional fast-Fourier transform (bottom panel, scale bar 2\,Å$^{-1}$) recorded on the Sb-terminated surface of Cs(V$_{0.95}$Ti$_{0.05}$)$_3$Sb$_5$. The kagome lattice structure is schematically indicated. (b)-(f), shown are two-dimensional maps (top, $I=4\,$nA, $V_{\rm m}=1\,$mV, scale bar 1\,nm) and corresponding two-dimensional fast-Fourier transforms (bottom panel, scale bar 2\,Å$^{-1}$) of the $dI/dV$ amplitude recorded in the field of view of panel a at $T=14\,$K and $V_{\rm B}=50,\,10,\,0,\,-10,\text{and}\,-30\,$mV, respectively. (g) and (h), shown are magnifications of the STM topography and $dI/dV$ map in panels a and c (white boxes), respectively. The kagome lattice (white lines) and $f-$wave charge order pattern (shaded triangles in panel h) are shown as an overlay to the data (scale bar 5\,Å). (i) Shown are $dI/dV$ spectra recorded at the center of the kagome lattice (black curve), on the triangle with high $dI/dV$ amplitude (blue curve), and on the triangle with low $dI/dV$ amplitude (red curve) ($V_{\rm B}=30\,$mV, $I=2\,$nA, $V_{\rm m}=1\,$mV, $T=14\,$K). The corresponding STM tip positions are indicated by identically colored dots in panels g and h. (j) Shown is a series of $dI/dV$ spectra recorded along the yellow line in panels a and c plotted as a function of the STM tip position and bias voltage ($V_{\rm B}=30\,$mV, $I=2\,$nA, $V_{\rm m}=1\,$mV, $T=14\,$K).}
    \label{fig:fig2}
\end{figure}

Charge order with ${\bf q}=0$ and finite angular momentum is characterized by spatial modulations of the charge density within a unit cell. To facilitate the detection of such subtle features in STM measurements, we first focus on titanium (Ti) doped CsV$_3$Sb$_5$, where effective hole doping of the electronic band structure suppresses the primary and secondary $2a\times2a$ and $4a\times1a$ CDWs~\cite{liu2023doping, pokharel2025evolution}. We have performed topographic and spectroscopic STM experiments on the surface of cleaved bulk crystals of Cs(V$_{0.95}$Ti$_{0.05}$)$_3$Sb$_5$ with $5\%$ Ti-doping (see Methods section) at different temperatures between 4\,K and 20\,K. In Fig.~\ref{fig:fig2}, we first focus on data recorded at an intermediate temperature $T=14\,$K. A typical STM topography and corresponding 2D-fast-Fourier-transform (2D-FFT) recorded on the sample surface are shown in Fig.~\ref{fig:fig2}(a). Apart from a Ti-doping induced smooth variations of the background signal, the STM topography is characterized by the honeycomb lattice of the topmost Sb layer~\cite{zhao2021cascade}. A closer inspection of the 2D-FFT only reveals the presence of six scattering vectors ${\bf q}_{\rm Bragg}$ with length ${q}_{\rm Bragg}=4\pi/\sqrt{3}a$, which result from periodic topographic modulations of the atomic lattice with lattice constant $a$, but no other scattering vectors. This is consistent with the absence of periodic lattice distortions induced by the $2a\times2a$ and $4a\times1a$ CDWs in real space~\cite{yang2022titanium}.

We conducted spectroscopic imaging experiments near the Fermi energy to examine the possible presence of $l>1$ charge orders in the low energy electronic structure of Cs(V$_{0.95}$Ti$_{0.05}$)$_3$Sb$_5$. Figure~\ref{fig:fig2}, (b)-(f) presents $dI/dV$ maps recorded at various bias voltages $V_{\rm B}$. Away from the Fermi energy ($V_{\rm B} = 50\,$mV), the $dI/dV$ map simply mirrors the apparent honeycomb lattice detected in the STM topography, and its 2D-FFT shows only the six Bragg peaks, ${\bf q}_{\rm Bragg}$. In stark contrast, a pronounced real-space modulation of the $dI/dV$ amplitude emerges in the maps near the Fermi energy ($|V_{\rm B}| \leq 10\,$mV). The 2D-FFT of this pattern reveals new scattering vectors appearing exclusively at wavevectors larger than the Bragg peaks (${\bf q} > {\bf q}_{\rm Bragg}$). The absence of any new peaks at ${\bf q} < {\bf q}_{\rm Bragg}$ confirms that this $dI/dV$ pattern preserves translational symmetry, corresponding to a modulation of the electronic structure within a single unit cell.

To examine the symmetry properties of this real space $dI/dV$ pattern, we plot a magnified view of the STM topography and $dI/dV$ map recorded at $V_{B}=10\,$mV in Fig.~\ref{fig:fig2}, (g) and (h), respectively. We also schematically overlay the kagome lattice (white lines) to register the $dI/dV$ modulations to the underlying crystal lattice. A close inspection of the $dI/dV$ map in Fig.~\ref{fig:fig2}(h) shows that the real space $dI/dV$ pattern is characterized by a periodic triangular structure, which is qualitatively distinct from the six-fold rotation symmetry ($C_{6z}$) of the underlying lattice structure. Specifically, we find that the $dI/dV$ pattern has large and small amplitude in neighboring triangles of the kagome lattice, exhibiting effective three-fold rotation  symmetry ($C_{3z}$). For comparison, we also overlay the $f-$wave charge order pattern on the kagome lattice, which breaks inversion ($\mathcal{P}$) and mirror-$xz$ ($\mathcal{M}_{xz}$) symmetries [Fig.~\ref{fig:fig1}(d)], and find that it closely aligns with the $dI/dV$ real space pattern. We note some local variations in the pattern's sharpness across the field of view seen in Fig.~\ref{fig:fig2}(c), such as a rounding of the triangle edges, which we attribute to disorder effects from the Ti dopant atoms. Indeed, as we show below, the agreement with the ideal $f$-wave model of Fig.~\ref{fig:fig1}(d) becomes nearly perfect in undoped CsV$_3$Sb$_5$ where such impurities are absent. Additional data sets that reproduce this observation are shown in Sec.~A of the suppl.~materials.

Moreover, the inversion-symmetry-breaking (ISB) nature of the $dI/dV$ pattern near Fermi energy is directly revealed in its 2D-FFT, shown in Fig.~\ref{fig:fig2}(c). While an FFT amplitude $F$ map is always inversion symmetric due to its inherent conjugate symmetry ($|F(\mathbf{q})| = |F(-\mathbf{q})|$), the detection of specific scattering vectors can still provide definitive evidence for ISB. The scattering vectors at ${\bf q}>{\bf q}_{\rm{Bragg}}$ jointly contribute to the fine structure of the spatial modulations of the $dI/dV$ amplitude; here, we specifically focus on the scattering vector ${\bf q}_{\mathrm{ISB}}$, which appears at twice the Bragg wavevector (${\bf q}_{\mathrm{ ISB}}=2\times{\bf q}_{\mathrm{ Bragg}}$) and along the same reciprocal space directions. The presence of this peak signifies a real-space modulation of the $dI/dV$ amplitude along the lattice vector directions with a period of half a lattice constant ($a/2$). As illustrated in the real-space schematic in Fig.~\ref{fig:fig2}(h), this $a/2$ modulation corresponds to an alternating pattern of high and low $dI/dV$ intensity on adjacent kagome triangles. This alternating pattern appearing on the kagome lattice explicitly breaks inversion symmetry, as seen in Fig.~\ref{fig:fig1}(d), thereby reducing $C_{6z}$ to $C_{3z}$ symmetry. This is the defining experimental signature of the $f$-wave charge order.

The inversion symmetry breaking $dI/dV$ pattern near the Fermi energy also manifests in spatial variations of the $dI/dV$ spectrum. In Fig.~\ref{fig:fig2}(i), we present $dI/dV$ spectra recorded at the kagome triangles with low (red line) and high amplitude (blue line), respectively, as well as in the center of the kagome honeycomb (black line). The spectrum recorded on the triangle with small $dI/dV$ amplitude is characterized by a wide U-shaped suppression of the $dI/dV$ amplitude near the Fermi energy appearing on top of a finite $dI/dV$ background. This spectrum can be distinguished from the spectrum recorded on the other triangle and honeycomb center, which are identical. We further map out these spectral modifications in real space and record a series of $dI/dV$ spectra along a line that intersects consecutive triangles with small and large $dI/dV$ amplitude (yellow line in Fig.~\ref{fig:fig2}(c)). The result of this spectroscopic line-cut is shown in Fig.~\ref{fig:fig2}(j). Consistent with the results of the $dI/dV$ map presented in panel c, the $dI/dV$ spectra exhibit a periodic modulation in real space, and the U-shaped $dI/dV$ spectra can be recorded on each triangle with low $dI/dV$ amplitude, as indicated by white arrows. Hence, the results of these spectroscopic measurements indicate that the real space modulation of the $dI/dV$ amplitude seen in Fig.~\ref{fig:fig2}(h) is related to the distinct U-shaped spectral gap appearing near the Fermi energy. Together, the results of our spectroscopic STM measurements suggest the presence of an inversion symmetry-breaking charge order, whose symmetry properties are consistent with a ${\bf q}=0$ $f-$wave ($l=3$) charge order shown in Fig.~\ref{fig:fig1}(d).

\subsection{Experimental characterization of a mirror and inversion symmetry breaking $q=0$ charge order}

\begin{figure}[H]
    \centering
    \includegraphics[width=1\linewidth]{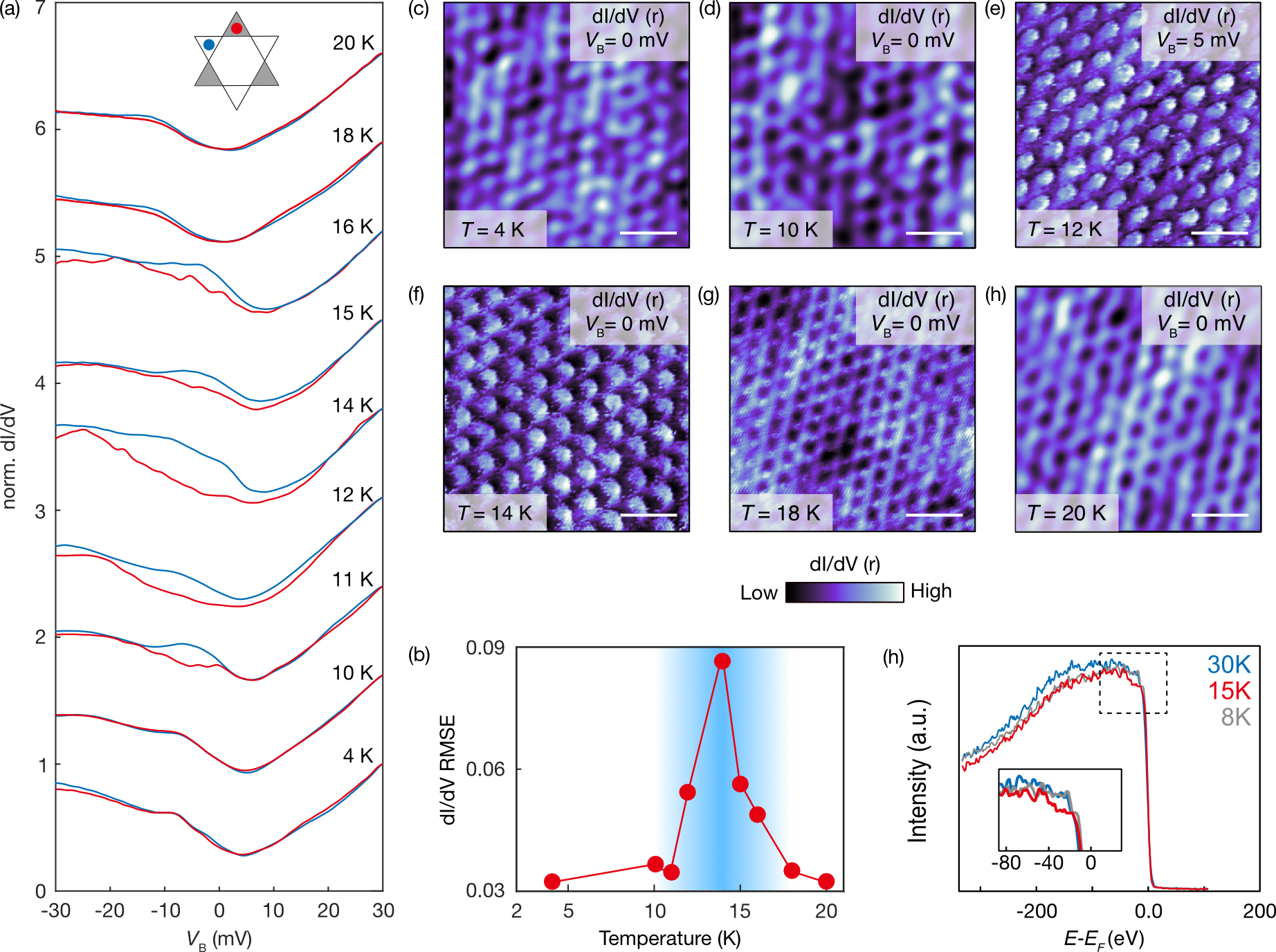}
    \captionsetup{singlelinecheck=off}
    \caption{\textbf{Intervening nature of inversion symmetry-breaking charge order detected in temperature-dependent measurements.} (a) Shown are a set of $dI/dV$ spectra acquired at the kagome triangle with high (blue line) and low (red line) $dI/dV$ amplitude recorded at different indicated temperatures ($V_{\rm B}=30\,$mV, $I=2\,$nA, $V_{\rm m}=1\,$mV). The STM tip position is schematically indicated in the inset. (b) Shown is the relative deviation between $dI/dV$ spectra recorded at the kagome triangle with high and low $dI/dV$ amplitude at different temperatures, quantified as the root mean square error (RMSE) [see Sec.~B of suppl.~materials]. (c)-(g), shown are $dI/dV$ maps recorded at the Fermi energy ($V_{\rm B}=0\,$mV) at different indicated temperatures (scale bar 1\,nm). (h) Shown is the energy-dependent amplitude of the measured ARPES spectrum at the crystal momentum of the Dirac band crossing near the Fermi energy [black arrow marker in Fig.~\ref{fig:fig1}(i)] at different indicated temperatures [full data sets are presented in Sec.~C of the suppl.~materials].}
    \label{fig:fig3}
\end{figure}

To experimentally characterize this charge order, we performed spectroscopic STM measurements at different temperatures. A comparison of $dI/dV$ spectra from the two distinct triangle types shown in Fig.~\ref{fig:fig3}(a) reveals a non-monotonic temperature evolution, which we quantify by plotting the root-mean-square error (RMSE) between the two spectra versus temperature in Fig.~\ref{fig:fig3}(b) [see Sec.~B. of suppl.~materials for details]. The RMSE, and thus the charge order itself, is negligible below 10\,K and above 18\,K. It emerges upon cooling below 18\,K, reaches a maximum intensity around 14\,K, and then abruptly disappears below 10\,K. This peak in the order's strength at 14\,K corresponds to the point where the distinction between the spectra is maximal, with one triangle type exhibiting a clear U-shaped suppression of the $dI/dV$ spectrum near Fermi energy [Fig.~\ref{fig:fig2}(a)]. The real-space $dI/dV$ patterns at representative temperatures confirm this non-monotonic evolution, though they reveal a more complex spatial evolution near the onset and disappearance of the order [Figs.~\ref{fig:fig3},~(c)-(g)].

Temperature-dependent ARPES measurements corroborate the spectral weight suppression near the Fermi energy seen in STM. Figure~\ref{fig:fig3}(h) shows normalized energy-dependent ARPES spectra recorded near the \mbox{$\Gamma-K$} Dirac crossing [$k_{x}\approx0.5\,\text{\AA}^{-1}$, see red arrow marker in Fig.~\ref{fig:fig1}(i)] at 8, 15, and 30\,K [full data are shown in Sec.~C of the suppl.~materials]. A suppression of spectral weight near the Fermi energy is discernible only in the 15\,K spectrum, while the 8\,K and 30\,K spectra are identical in this region. The energy scale of this effect, $\approx30\,$meV as seen in the inset of Fig.~\ref{fig:fig3}(h), is consistent with our STM results [Fig.~\ref{fig:fig2}(i)]. The effect appears weak in ARPES because it is a spatially-averaging probe, whereas STM shows the charge order is spatially inhomogeneous with the U-shaped $dI/dV$ spectrum appearing only at a quarter of the surface area [Fig.~\ref{fig:fig2}(j)]. The combined STM and ARPES data therefore indicate that the charge order is an intervening phase, existing only in a narrow temperature window between approximately 10 and 18\,K.

\begin{figure}[H]
    \centering
    \includegraphics[width=1\linewidth]{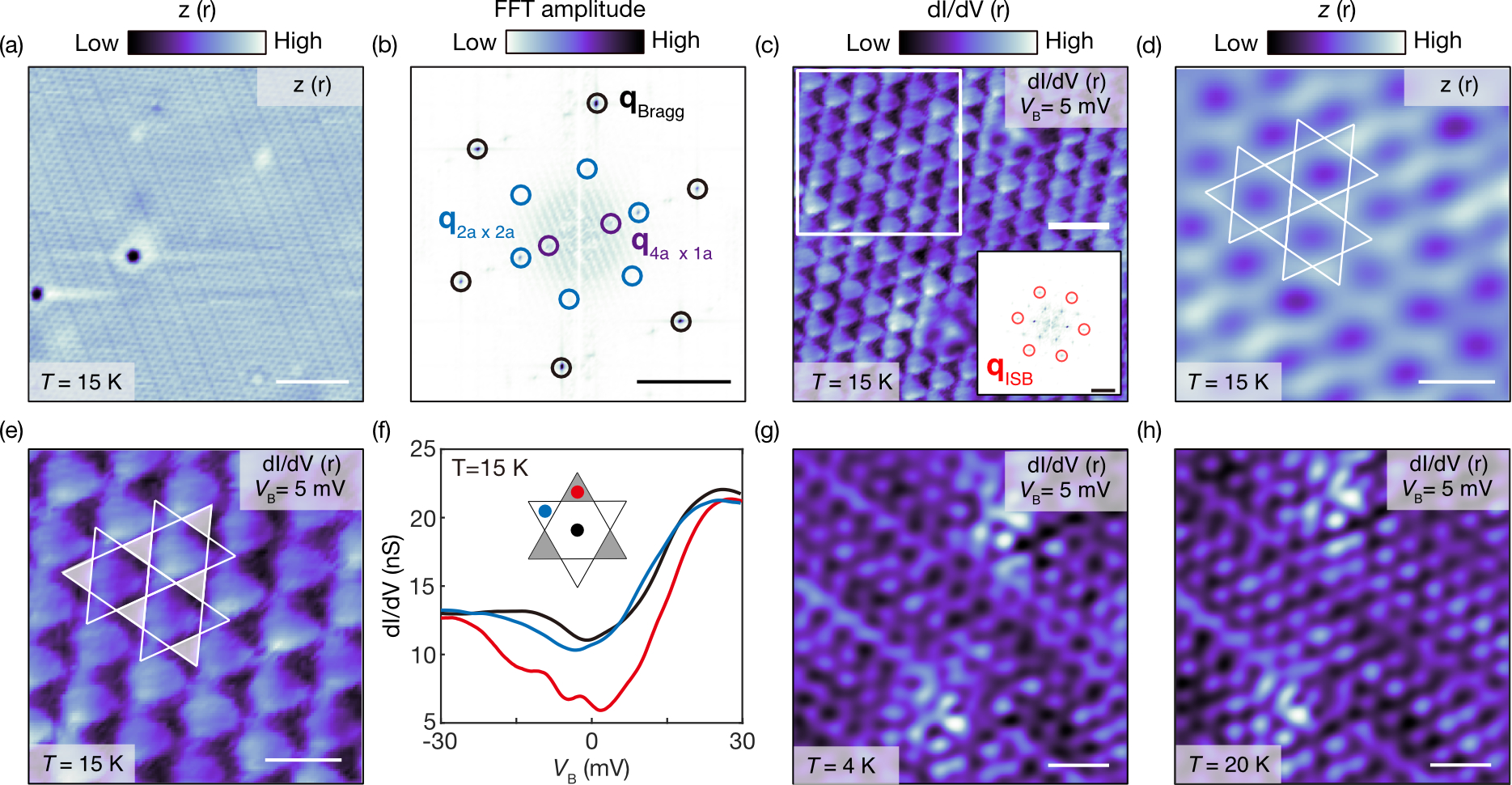}
    \captionsetup{singlelinecheck=off}
    \caption{\textbf{Insensitivity of inversion symmetry-breaking charge order to titanium doping level.} (a) and (b), Shown are an STM topography (scale bar 5\,nm) and the corresponding 2D-FFT (scale bar 1\,Å$^{-1}$) of the Sb-terminated surface of CsV$_{3}$Sb$_5$, respectively ($V_{\rm B}=-0.5\,$V, $I=100\,$pA, $T=15\,$K). The Bragg vectors ${\bf q}_{\text{Bragg}}$, and scattering vectors ${\bf q}_{2a\times2a}$ and ${\bf q}_{4a\times1a}$ of the $2a\times2a$ and $4a\times1a$ CDWs are indicated. (c) Shown is a $dI/dV$ map recorded near the Fermi energy ($V_{\rm B}=5\,$mV) at $T=15\,$K on the Sb-termined CsV$_{3}$Sb$_5$ surface ($V_{\rm m}=1\,$mV, $I=4\,$nA). The inset shows the corresponding 2D-FFT (scale bar 2\,Å$^{-1}$). The inversion symmetry breaking (ISB) scattering vector ${\bf q}_{{\rm ISB}}$ is highlighted by red circles. (d) and (e), shown are magnifications (white box in panel c) of the STM topography and the $dI/dV$ map in panel c. The kagome lattice (white lines) and $f-$wave charge order pattern (shaded triangles in panel k) are shown as an overlay to the data (scale bar 5.5\,Å). (f) Shown are $dI/dV$ spectra recorded at the center of the kagome lattice (black curve), on the triangle with high $dI/dV$ amplitude (blue curve), and on the triangle with low $dI/dV$ amplitude (red curve) ($V_{\rm B}=100\,$mV, $I=3\,$nA, $V_{\rm m}=1\,$mV, $T=15\,$K). The corresponding STM tip positions are schematically indicated by identically colored dots in the inset. All scale bars (unless otherwise noted) in topography and $dI/dV$ maps correspond to 1\,nm. (g) and (h), Shown are $dI/dV$ maps recorded in the sample field of view as panel c near the Fermi energy ($V_{\rm B}=5\,$mV) at $T=4\,$K and $T=20\,$K, respectively ($V_{\rm B}=50\,$mV, $V_{\rm m}=1\,$mV, $I=4\,$nA).}
    \label{fig:fig4}
\end{figure}

To test the robustness of this charge order against changes in chemical potential, we performed STM measurements on pristine, undoped CsV$_3$Sb$_5$ at 15\,K. This is significant because previous work has shown that the conventional $2a\times2a$ CDW is suppressed by doping~\cite{yang2022titanium, pokharel2025evolution}. The STM topography, shown in Fig.~\ref{fig:fig4}(a), is characterized by the typical periodic lattice distortions of the topmost Sb layer due to the presence of the $2a\times2a$ and unidirectional $4a\times1a$ CDWs~\cite{zhao2021cascade}. Similarly, the 2D-FFT features the scattering vector of the atomic lattice ${\bf q}_{\text{Bragg}}$ alongside the scattering vectors, ${\bf q}_{2a\times2a}$ and ${\bf q}_{4a\times1a}$ of the CDWs [Fig.~\ref{fig:fig4}(b)]~\cite{zhao2021cascade, jiang2021unconventional}. In stark contrast, high-resolution $dI/dV$ maps recorded over a smaller field of view on the same surface reveal the clear triangular, ${\bf q}=0$ modulation of the $dI/dV$ amplitude characteristic of the $f$-wave order, identical to that seen on the doped samples [Fig.~\ref{fig:fig4}(c)]. Notably, this pattern appears even more uniform and well-developed than in the doped samples, likely due to the absence of disorder from Ti impurities. A detailed symmetry analysis of this pattern [Figs.~\ref{fig:fig4}(e)] and the observation of the ${\bf q}_{\rm ISB}$ peak in its 2D-FFT [inset, Fig.~\ref{fig:fig4}(c)] confirm it has the symmetry properties of an inversion-breaking $f$-wave charge order seen in Fig.~\ref{fig:fig1}(d). The presence of this charge order in both doped and undoped samples demonstrates its robustness and independence from conventional CDWs previously detected in this material~\cite{zhao2021cascade, jiang2021unconventional}.

Further spectroscopic characterization of this charge order in CsV$_3$Sb$_5$ reveals properties consistent with our findings on the doped compound. Spatially-resolved $dI/dV$ point spectra confirm the U-shaped suppression near Fermi energy on only one sub-set of kagome triangles [Fig.~\ref{fig:fig4}(f)], and bias-dependent mapping of the $dI/dV$ amplitude shows this pattern exists only near the Fermi energy [see Sec.~A.2 of suppl.~materials]. Furthermore, the order exhibits the same non-monotonic temperature dependence, appearing only within an intermediate temperature window [Figs.~\ref{fig:fig3},~(g) and (h)]. Finally, the analysis of large-area $dI/dV$ maps recorded near the Fermi energy at $T=15\,$K shows that this charge order coexists with the underlying $2a\times2a$ CDW modulation [Sec.~D of the suppl.~materials].

\subsection{Theoretical understanding of $f$-wave charge bond order in CsV$_3$Sb$_5$}
Our experimental study demonstrates the observation of an inversion symmetry-breaking $dI/dV$ pattern with ${\bf q}=0$ near Fermi energy concomitant with the suppression of the ARPES intensity near the Dirac crossing on the $\Gamma-K$ line in the Brillouin zone at temperatures around 14\,K. In the following, based on symmetry arguments and model calculations, we will argue that this observation can be naturally explained by the presence of an $f$-wave charge bond order (CBO) with ${\bf q}=0$ promoted by the unique sublattice texture of the kagome lattice band structure~\cite{zhan2026loop}. We will then address the intervening nature of this phase detected in temperature-dependent measurements and discuss its place in the phase diagram of correlated phases of CsV$_3$Sb$_5$.

To conceptually understand the role of different parts of the Fermi surface for the stabilization of an $f$-wave CBO, we first employ a minimal two-orbital kagome model (see Methods section for model details). This model qualitatively reproduces the essential low-energy band structure of CsV$_3$Sb$_5$, notably the van Hove singularity at $M$ and the Dirac cone along the $\Gamma-K$ direction, which is composed of $d_{xz,\,yz}$ vanadium bands, as seen in Fig.~\ref{fig:fig1}(f)~\cite{zeng_electronic_2025}. Within this model, we construct a ${\bf q}=0$ $f$-wave CBO that has the required odd parity ($B_{2u}$) symmetry by modulating nearest-neighbor hoppings shown in Fig.~\ref{fig:fig1}(d). Crucially, this order parameter gaps the states along the $\Gamma-K$ direction but breaks no symmetries along $\Gamma-M$. Consequently, it selectively opens a spectral gap at the Dirac cone while leaving the states near the $M-$point unaffected [Fig.~\ref{fig:fig1}(f)]. A susceptibility analysis, detailed in the Sec.~E of the suppl.~materials, confirms that the electronic states at the Dirac cone are indeed highly susceptible to this specific $f$-wave CBO owing to their mixed sub-lattice texture, unlike the states at the $M-$point, which have pure sub-lattice character~\cite{zhan2026loop}.

\begin{figure}[H]
    \centering
    \includegraphics[width=1\linewidth]{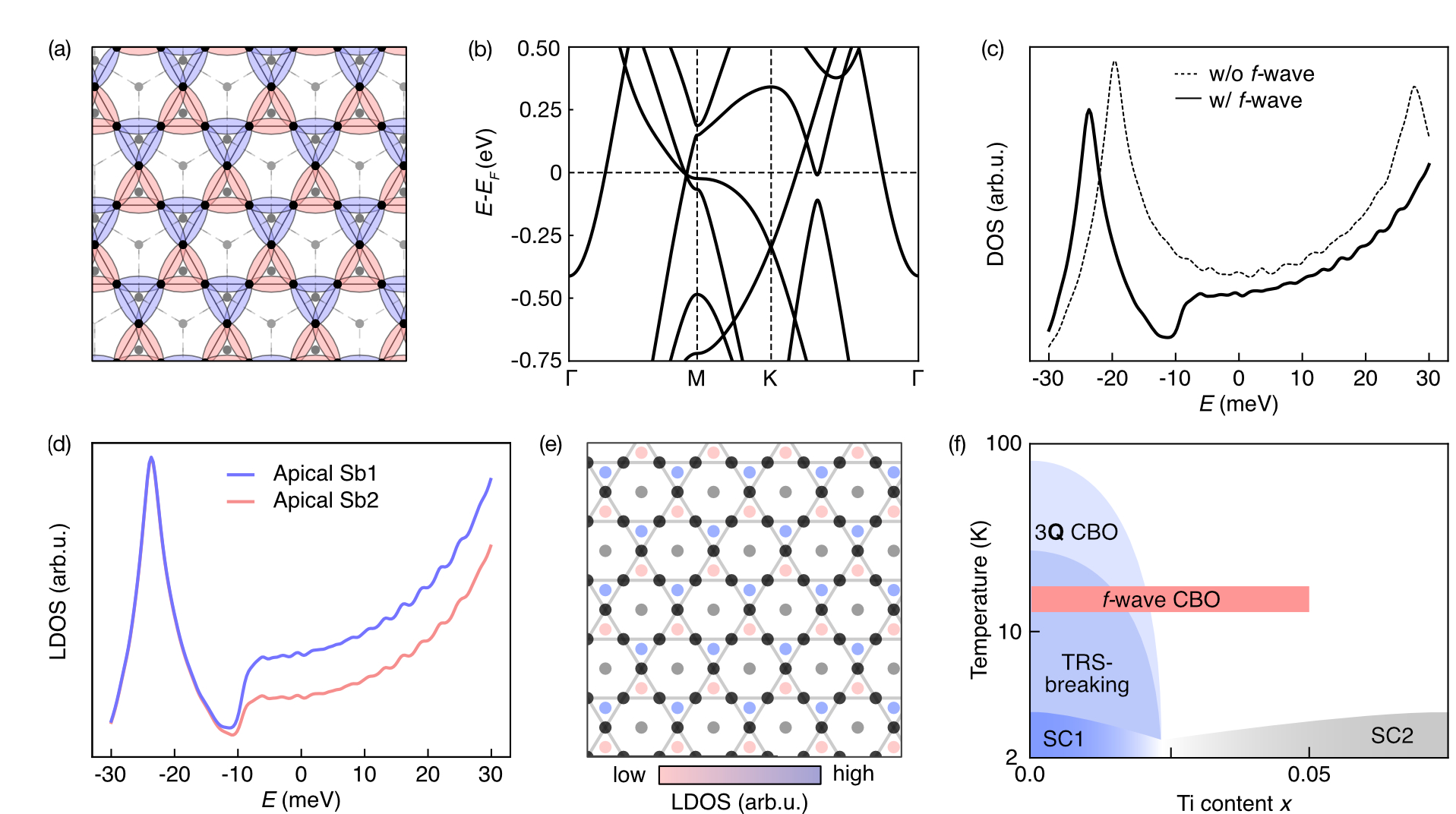}
    \caption{\textbf{Odd-parity charge order with $f-$wave form factor in CsV$_{3}$Sb$_5$.} (a) Shown is the kagome lattice of CsV$_{3}$Sb$_5$, including Sb atoms (grey dots) and V atoms (black dots) with the $f-$wave charge bond order (CBO) overlaid in color. The red and blue bonds denote increased/decreased hopping parameters, respectively as schematically shown in Fig.~\ref{fig:fig1}(d). The $C_{6z}$ symmetry is broken down to $C_{3z}$, and inversion symmetry is also broken. (b) Shown is the band structure of CsV$_{3}$Sb$_5$ in the presence of the $f$-wave charge bond order, which gaps the Dirac crossing near Fermi energy along the $\Gamma$-$K$ line. (c) Shown is the total density of states (DOS) with and without $f$-wave CBO with amplitude $\Delta=20\,$meV. (d) Shown is the calculated local density of states (LDOS) at the sites of the two types of apical Sb sitting on top of the kagome triangles. (e) Shown is the calculated real-space resolved LDOS in the presence of $f$-wave CBO at $E=-10\,$meV overlaid to apical Sb atoms. Also shown are the vanadium kagome atoms (black dots) and the in-plane Sb atoms (grey dots). (f) Shown is a simplified phase diagram of standard charge order and superconducting (SC) phases that are well established in the literature, with the addition of the $f$-wave CBO that we have identified in this work. }
    \label{fig:fig5}
\end{figure}

Next, we present results from realistic 30-band tight-binding model calculations to account for the experimental phenomenology observed in STM and ARPES measurements at a more quantitative level. To construct a model for $f$-wave CBO, we first consider the electronic states in the absence of any charge order. For this, we apply a Wannier orbital-based tight-binding model derived from density functional theory, which accurately reproduces the band structure of CsV$_3$Sb$_5$ in a wide range around the Fermi energy, as seen in Fig.~\ref{fig:fig1}(h)~\cite{kang2022twofold} [also see Sec.~F of the suppl.~materials]. The model includes $d$-orbitals on the V sites, as well as $p$-orbitals on the Sb sites. Importantly, it also includes the apical Sb sites, which are above/below the center of the kagome triangles, as illustrated in Fig.~\ref{fig:fig5}(a). These sites constitute the top-most Sb-layer of our sample, where symmetry breaking is detected in measurements of the local density of states with the STM. 

To add the $f-$wave CBO, we have performed a group theory classification of all possible ${\bf q}=0$ bond orders that occur intra-unit cell, i.e., that do not break translation symmetry with respect to the original, undistorted, kagome lattice. A full classification is shown in Sec.~G of the suppl.~materials, but the one we use henceforth is the $f$-wave pattern, which more precisely lies in the $B_{2u}$ irreducible representation of $D_{6h}$. The important takeaway is the $B_{2u}$ breaks inversion, the mirror $M_{zx}: y \to -y$, and the sixfold rotation $C_{6z}$ down to $C_{3z}$. It preserves however $M_{yz}: x \to -x$. Note that the $x,y$ axes are defined in Fig.~\ref{fig:fig1} (a). A schematic of this CBO superimposed to the lattice structure of CsV$_3$Sb$_5$ is shown in Fig.~\ref{fig:fig5}(a), wherein the red/blue bonds denote increased/decreased hopping strength, respectively. The details of how we introduce this $f$-wave order parameter to the model are explained in Sec.~H of the suppl.~materials and in the following, we will focus on the observable consequences.

Because the $\Gamma$--$K$ line is invariant under $k_y\to-k_y$ and $k_z\to -k_z$, electron wave functions on this line may be classified according to their mirror eigenvalues under $M_{zx}$ and $M_{xy}$. Furthermore, we know that both bands participating in the crossing are odd under $M_{xy}$ \cite{zeng_electronic_2025}. Hence, the symmetry operation which allows this crossing is $M_{zx}$, under which the two wave functions have distinct eigenvalues. Breaking this mirror symmetry by the $f-$wave order parameter, therefore, requires the two states to experience a level repulsion. This leads to a gapping of the Dirac crossing on the $\Gamma$--$K$ line, as seen in Fig.~\ref{fig:fig5}(b). This understanding is consistent with our observation of a suppressed ARPES intensity in our measurements at $T=15\,$K [Fig.~\ref{fig:fig3}(h)].

Let us now consider the experimental observation of the ${\bf q}=0$ inversion-symmetry breaking $dI/dV$ pattern and the U-shaped suppression of the $dI/dV$ amplitude near Fermi energy detected in our STM measurements. We first compare the calculated density of states (DOS) in the presence and absence of the $f-$wave CBO in Fig.~\ref{fig:fig5}(c). We find that the gapping of the Dirac cone by this state seen in Fig.~\ref{fig:fig5}(b) results in a suppression of the DOS amplitude slightly below the Fermi level, which is in agreement with our experimental observation in Fig.~\ref{fig:fig2}(i). Note that the large peak in the DOS spectrum at $\approx-20\,$meV is contributed by the van-Hove singularity at the $M-$point and is unrelated to electronic states of the Dirac point along $\Gamma-K$. Furthermore, the $f-$wave CBO breaks the $\mathcal{M}_{zx}$ symmetry, which relates the two different apical sites within the Sb honeycomb layer, which in our STM measurements is the surface layer. Consequently, the sublattice-resolved plot of the calculated local density of states (LDOS) in Figs.~\ref{fig:fig5}(d) shows a clear splitting between the two apical Sb sites due to this symmetry breaking, consistent with our experimental observations [Fig.~\ref{fig:fig1}(i)]. The difference between the LDOS spectra at the two apical Sb sites is also reflected in the real-space-resolved plot of the LDOS amplitude, shown in Fig.~\ref{fig:fig5}(e). The LDOS pattern breaks the $\mathcal{P}$ and $\mathcal{M}_{xz}$ symmetries of the kagome lattice and exhibits a $C_{3z}$-symmetry, which reproduces our experimental observations. The excellent agreement between the results of our realistic tight-binding model calculation and the salient experimental observations of our STM measurements provides compelling support for the observation of an odd-parity $f-$wave CBO on the surface of CsV$_3$Sb$_5$.

\subsection{Discussion and Conclusion}
We now turn to a brief discussion on the possible source of $f$-wave CBO, and the contrast with the primary Star-of-David/$3\bf Q$ CBO appearing around 90\,K, which is associated with the $2a\times2a$ CDW [see phase diagram of Fig.~\ref{fig:fig5}(f)]. Indeed, the vast majority of theoretical works on charge order in CsV$_3$Sb$_5$ have focused on CBOs concerning electrons near the $M$-points~\cite{kiesel2013unconventional, feng2021chiral, denner2021analysis, park2021electronic, fu2025exotic}, because of the van Hove singularities in close proximity to the Fermi level. At lower temperatures, various experimental studies have reported the detection of additional rotation and time-reversal symmetry-breaking (TRS) transitions inside this primary CDW phase at lower temperatures~\cite{nie2022charge, guo2022switchable, khasanov2022time, gui2025probing, wang2025long, fernandes2025loop}, also attributed to $M$-point electrons. On the other hand, our susceptibility analysis shows that the $f-$wave CBO discovered in this study has vanishing weight near the $M-$point and is primarily stabilized by the Dirac cone on the $\Gamma-K$ line, which is in agreement with results from our ARPES measurements. Note that this vanishing weight at the $M$-point is a consequence of the odd parity of the $f$-wave CBO combined with the symmetry properties and sublattice composition of the wave functions at the van Hove singularities~\cite{fu2025exotic,zhan2026loop}. Hence, our combined results suggest that the $f-$wave and $3\bf Q$ CBOs are promoted by different parts of the Fermi surface. This understanding would also account for their contrasting response to Ti-doping seen in our measurements where $f-$wave CBO is still present but $3\bf Q$ CBO is suppressed at 5$\,\%$ doping. While the underlying mechanism driving the formation of this $f-$wave phase remains an open question for future studies (i.e. electron-electron driven vs. electron-phonon driven~\cite{fernandes2014drives}), a renormalized susceptibility analysis indicates that nearest-neighbor Coulomb repulsion strongly enhances the tendency toward formation of odd-parity charge orders [see Sec.~E of the suppl.~materials].

In summary, combining high-resolution spectroscopic mapping with the STM with comprehensive theoretical modeling, we have presented the first direct evidence for an inversion symmetry-breaking $f$-wave charge order in the kagome metal CsV$_3$Sb$_5$. Our analysis identifies this state as a ${\bf q}=0$ bond order stabilized by the gapping of a Dirac point. This work therefore elevates odd-parity electronic orders from a theoretical proposal to an experimentally realized phase of matter. Perhaps our most intriguing finding, however, is that this $f-$wave CBO is itself an intervening phase, vanishing below 10\,K into a new state that leaves no definitive symmetry-breaking signature in our STM measurements, which are sensitive to the local electronic density of states. For example, $dI/dV$ spectra near Fermi energy recorded above (20\,K) and below (4\,K) the $f-$wave phase are nearly identical and show no real-space variation~[Fig.~\ref{fig:fig3}(a)]. This points to the emergence of a 'hidden' order below 10\,K. Our discovery thus brings several fundamental challenges to the forefront. A key task will be to understand the interplay between the odd-parity order arising from electronic states near the $K$-point and the primary $3\bf Q$ CBO~\cite{jiang2021unconventional, zhao2021cascade, mielke2022time, xu2022three} and symmetry-breaking orders~\cite{nie2022charge, guo2022switchable, khasanov2022time, gui2025probing, wang2025long} originating near the $M$-points. Furthermore, determining the interplay between the intervening odd-parity order we have discovered, this emergent hidden state and the superconducting phases detected at low temperatures~\cite{ortiz2019new, chen2021roton, yang2022titanium, deng2024chiral}, as illustrated in Fig.~\ref{fig:fig5}(f), will be key to constructing a complete model of the correlated phase diagram in the kagome metals.

\clearpage
\bibliography{bibliography}

\clearpage
\section{Figures}
\clearpage

\setcounter{figure}{0}
\makeatletter 
\renewcommand{\figurename}{Extended Data Figure}
\makeatother

\clearpage
\section{Methods}

\subsubsection{Tight-binding model calculations}
We construct an effective two-orbital model to describe the low-energy electronic structures. The Hamiltonian reads,
\begin{equation}
    H_{\text{2}} =  \sum_{\langle i,j \rangle,\alpha\beta\sigma} t^{\alpha\beta}c_{i\alpha\sigma}^\dagger c_{j\beta\sigma} +\sum_{\langle\langle i,j \rangle \rangle,\alpha\beta\sigma} s^{\alpha\beta}c_{i\alpha\sigma}^\dagger c_{j\beta\sigma}  -  \sum_{i,\sigma} \mu_\alpha c_{i\alpha\sigma}^\dagger c_{i\alpha\sigma}.
\end{equation}
Here $c^\dag_{i\alpha\sigma}$($c_{i\alpha\sigma}$) creates (annihilates) an electron at the lattice site $i$ in orbital $\alpha$. $\alpha=1(2)$ denotes the $d_{xz}$($d_{yz}$) orbital. The hopping parameters are given in units of eV as follows: $t^{11}=-0.4$, $t^{22}=0.28$,  $t^{12}=0.45$, $s^{11}=-0.15$, $s^{22}=-0.06$,  $s^{12}=0.08$, $\mu_1=0.14$, $\mu_2=-0.12$, $t^{12}=-t^{21}$ and $s^{12}=-s^{21}$. 

The odd-parity $f$-wave order has an angular moment of $l=3$. It preserves the translational symmetry and the mirror symmetry $\mathcal{M}_{yz}$, but breaks the mirror symmetry $\mathcal{M}_{xz}$ (corresponding to a mirror reflection along the $\Gamma-K$ line). Within the two-orbital model, this order can be decomposed into intra- and inter-orbital contributions, with the corresponding Hamiltonian given by 
\begin{eqnarray}
    H_{f,\text{intra}} & =& \Delta_0^\text{intra}  \sum_{\bm{R}\alpha} \sum_{a,\epsilon_{abc}=1}  (c^\dagger_{\bm{R}a, \alpha \sigma} c_{\bm{R}b ,\alpha,\sigma} -c^\dagger_{\bm{R}a ,\alpha \sigma} c_{\bm{R}-\bm{l}_c b, \alpha\sigma}), \\
    H_{f,\text{inter}} & = & \Delta_0^\text{inter}  \sum_{\bm{R}\alpha\beta} \sum_{a,\epsilon_{abc}=1} \left[  (c^\dagger_{\bm{R}a, \alpha\sigma} c_{\bm{R}b ,\beta\sigma} -c^\dagger_{\bm{R}a ,\alpha \sigma} c_{\bm{R}-\bm{l}_c b, \beta\sigma})-(c^\dagger_{\bm{R}a, \beta \sigma} c_{\bm{R}b ,\alpha \sigma} -c^\dagger_{\bm{R}a,\beta \sigma} c_{\bm{R}-\bm{l}_cb , \alpha\sigma})\right].
\end{eqnarray}
Here $\bm{R}a$ labels the lattice site $i$, $\bm{R}$ is the unitcell coordinate and $a,b,c\in {1,2,3}$ are the sublattice indices in the kagome lattice.  The vector $\bm{l}_{c}$ is the lattice vector parallel to the bond connecting $a$ and $b$ sublattices, with Levi-civta symbol satisfying $\epsilon_{abc}=1$. Due to the inter-band contribution to the susceptibility from the band crossing along the $\Gamma-K$, the $f$-wave order can emerge. Once established, the breaking $\mathcal{M}_{xz}$ symmetry opens a gap at this band crossing, as shown in the Fig 1(i) using the parameters  $\Delta^\text{intra}_0=0$ and $\Delta^\text{inter}_0=0.2t^{\alpha\beta}$.

\subsubsection{Synthesis of (Ti-doped) CsV$_3$Sb$_5$ crystals}

Single crystals of CsV$_3$Sb$_5$ and Cs(V$_{0.95}$Ti$_{0.05}$)$_3$Sb$_5$ were grown via a conventional flux-based growth technique. Vanadium powder purchased from the Sigma-Aldrich was cleaned using a mixture of isopropyl alcohol and Hydrochloric acid to remove residual oxides. Cs (liquid, Alfa 99.98\%), V (powder, Sigma 99.9\%), Ti (powder, Alfa 99.9\%), and Sb (shot, Alfa 99.999\%)  were loaded inside a milling vial with the required stoichiometry and then sealed in an Argon-atmosphere. To grow the single crystals of Cs(V$_{1-x}$Ti$_x$)$_3$Sb$_5$ ($x$=0, 0.05), elemental stoichiometry, respectively, taken as Cs$_{20}$V$_{15-x}$Ti$_x$Sb$_{120}$ ($x$=0, 3) and then milled for about an hour. After milling, the powders were poured into the alumina crucibles and sealed inside a stainless steel tube. The samples were heated at 1000$^\circ$ C for 10 h and then cooled to 900$^\circ$ C at 25$^\circ$ C/hr. Below 900$^\circ$ C, the samples were cooled at 1$^\circ$ C/hr to 500$^\circ$ C. Once the growth period was over, the plate-like single crystals were separated gently from the flux and then cleaned with ethanol.  

\subsubsection{Angle-resolved photoemission spectroscopy (ARPES) measurements}
ARPES measurements were performed using a lab-based system with a Helium lamp (He-I, $E=21.2\,$eV) and a Scienta Omicron DA30-L electron analyzer. The Cs(V$_{0.95}$Ti$_{0.05}$)$_3$Sb$_5$ samples were cleaved and in-situ transferred to the measurement system at a base pressure better than $5\times10^{-11}\,$Torr. The energy resolution of the ARPES measurement is better than 10\,meV at $T=15\,$K.

\subsubsection{Scanning Tunneling Microscopy (STM) Measurements}

The undoped and titanium-doped CsV$_{3}$Sb$_5$ samples were cleaved after cooling down to a temperature $T=4\,$K inside an ultra-high vacuum (UHV) chamber with a base pressure of $p\approx1.4 \times 10^{-10}\,$mbar. Several crystals of doped and undoped CsV$_{3}$Sb$_5$ were cleaved and the results presented in this manuscript were consistently observed. STM measurements were conducted using a home-built STM instrument under cryogenic (4\,K$\leq T\leq20\,$K) and UHV ($p\approx1.4 \times 10^{-10}\,$mbar) conditions using a chemically etched tungsten tip. The tip was prepared on a Cu(111) surface through field emission and controlled indentation, as well as calibrated against the Cu(111) Shockley surface state before each set of measurements. Bias voltage ($V$) dependent differential conductance ($dI/dV$) spectra and maps were recorded using standard lock-in methods with a bias modulation $1\,\text{mV}\leq V_{\rm m}\leq10\,\text{mV}$ at a frequency $f=3.971\,$kHz, as indicated in the main text. The $dI/dV$ maps were recorded using multi-pass mode to avoid set-point effects. To enhance the clarity of the data presentation, high-frequency noise originating from mechanical vibrations coupling to the tip-sample junction was removed from the raw data and interpolation was performed to reduce pixelated appearance where applicable.


\clearpage
\section{Acknowledgments}

The authors appreciate valuable discussions with Adrian Po and J\"org Schmalian. This work was primarily supported by the Hong Kong Research Grant Council (Grant Nos.\,26304221, 16302422, 16302624, 16304525, C6033-22G, and AoE/P-604/25R-I awarded to BJ), the Croucher Foundation (Grant No.\,CIA22SC02 awarded to BJ, and the NSFC/RGC Collaborative Research Scheme (CRS\_CityU101/25). XW and SZ are supported by the National Key R\&D Program of China (Grants No. 2023YFA1407300 and No. 2022YFA1403800) and the National Natural Science Foundation of China (Grants No. 12574151, No. 12447103, No. 12447101, No. 12374153, and No. 12047503). DJS acknowledges support by the German Research Foundation (DFG) through CRC TRR 288 “Elasto-Q-Mat,” project A07. Work at the University of Minnesota was supported by the National Science Foundation through the University of Minnesota MRSEC under Award Number DMR-2011401. JM acknowledges support by the Hong Kong Research Grants Council (Grant No.21304023), the National Natural Science Foundation of China (12422405), and the NSFC/RGC Collaborative Research Scheme (CRS\_CityU101/25). SDW, ACS, and GP gratefully acknowledge support via the UC Santa Barbara NSF Quantum Foundry funded via the Q-AMASE-i program under award DMR-1906325.

\section{Author Contributions}

JZ, CC, and BJ conceived the project. JZ and CC carried out the STM measurements and analyzed the data. XW theoretically proposed the f-wave order and analyzed its physical consequences. RF carried out the two-orbital kagome lattice model and susceptibility calculations under the supervision of XW and SZ. DJS performed the symmetry analysis and realistic tight-binding model calculations assisted by LB. ZL, FY, and TG performed the ARPES measurements and related data analysis under the supervision of JM. GP and ACS grew the bulk crystals under the supervision of SW. All authors contributed to the discussion and manuscript, which was written by BJ.

\section{Competing Interest Declaration} The authors declare that they have no competing financial interest.

\section{Data Availability Statement} Replication data for this study can be accessed on Zenodo via the link XXX.

\end{document}